\documentclass[aps,pre,showkeys,showpacs,superscriptaddress,reprint,longbibliography,
nofootinbib,
floatfix]{revtex4-1}
\usepackage[utf8]{inputenc}
\usepackage[T1]{fontenc}
\usepackage{amsmath}
\usepackage{amsfonts}
\usepackage{graphicx}
\usepackage[colorlinks=true,hyperfootnotes=true,breaklinks=true]{hyperref}

\begin{document}
\title{Influence of a~range of~interaction among agents on~efficiency and~effectiveness of~knowledge transfer within an~organisation}

\author{Kamil Paradowski}
\affiliation{\href{http://www.agh.edu.pl/}{AGH University of Science and Technology},
\href{http://www.pacs.agh.edu.pl/}{Faculty of Physics and Applied Computer Science},\\
al. Mickiewicza 30, 30-059 Krakow, Poland}

\author{Agnieszka Kowalska-Stycze\'n}
\email{Agnieszka.Kowalska-Styczen@polsl.pl}
\affiliation{\href{http://www.polsl.pl/}{Silesian University of Technology},
Faculty of Organisation and Management,\\
ul. Roosevelta 26/28, 41-800 Zabrze, Poland}

\author{Krzysztof Malarz}
\homepage{http://home.agh.edu.pl/malarz/}
\email{malarz@agh.edu.pl}
\affiliation{\href{http://www.agh.edu.pl/}{AGH University of Science and Technology},
\href{http://www.pacs.agh.edu.pl/}{Faculty of Physics and Applied Computer Science},\\
al. Mickiewicza 30, 30-059 Krakow, Poland}

\date{\today}
\begin{abstract}
In this study we examined how the size of non-formal groups between organization members affect the transfer of knowledge in the context of the efficiency and effectiveness of this process.
To analyse the dynamics of the transfer of knowledge the cellular automata model was used.
The model is based on local interactions between members of the organization, that take place in the nearest neighbourhood.
These groups of close neighbours are represented by von Neumann’s neighbourhood (four nearest-neighbours) and Moore’s neighbourhood (four nearest-neighbours and four next-nearest neighbours) and complex neighbourhood (four nearest neighbours, four next-nearest neighbours and four next-next-neighbours).
The results of the simulation show the influence of the size of the neighbourhood on the efficiency of knowledge transfer.
\end{abstract}

\date{\today}



\maketitle

\section{Introduction}

Today in a fast-changing environment, knowledge is the dominant source of competitive advantage~\cite{Chen-2004,Lyles-1996,Tsai-2001}.
In literature, there are many definitions of knowledge.
For example, Applehans et al. \cite{Applehans-1999} define knowledge as information used to solve a problem.
Davenport and Prusak \cite{Davenport-2000} point out, that knowledge exists in people and is an inherent part of human complexity and unpredictability.
This is also confirmed by the Buckman of Bucman Labs study, which shows that 90\% of the knowledge in each organization is contained in ``people's heads'' \cite{Wah-1999}. 
The core process in an organization, where knowledge is present is a knowledge transfer.
Knowledge transfer in an organization is the process through which one person or group influences the experience of others \cite{Argote-2000}. 
Knowledge is created when people communicate and share knowledge, assimilate and apply what they have learned.
As knowledge transfer is essential for many organizational processes, including best practice transfer, product development and organizational survival \cite{Reagans-2003}, its effectiveness and efficiency are particularly important.

Knowledge and its distribution are strongly linked to social interactions.
Managers receive two-thirds of information and knowledge through face-to-face communication or telephone conversations, and only one-third come from documents \cite{Davenport-2000}.
The importance of social interaction, especially informal contacts in the knowledge transfer process, has been demonstrated by many authors \cite{Ingram-2000,Reagans-2001,Reagans-2003,Chen-2007,Tsai-2001}.
Therefore, in this article, knowledge transfer is understood as a common creative process in an organization, which is most often done informally, by sharing face-to-face knowledge, as in Ref. \cite{Girdauskiene-2012}.

Our goal is simulation and bottom-up approaches in modelling the transfer of knowledge, where local (bottom-up) relationships generate phenomena at a global level (that is, at the level of the whole organization).
The basic premise of the model, inspired by Reagans and McEvily \cite{Reagans-2003}, is to divide the knowledge transferred into a number of portions (chunks) of knowledge.
Informal contacts between members of the organization are represented in our research by different neighbourhoods size (four, eight or twelve elements), because groups of employees consist of a number of members.

The results presented here base on model of knowledge transfer within a small or a medium organisation~\cite{Kowalska-2017a-e} where agents send and recipe chunks of knowledge only when distances
\emph{i}) in space among agents
\emph{ii}) and in knowledge are small.
Removing the latter restriction leads to more efficient and more effective knowledge transfer~\cite{Kowalska-2017b-e}.

Here we would like to check if omitting the spatial restriction also may be helpful in spreading knowledge among agents in artificial organisation.
Similarly to our earlier approaches~\cite{Kowalska-2017a-e,Kowalska-2017b-e} we will predicate our discussion of the results on the computer simulations based on cellular automata (CA) technique.

\section{\label{S:model}Model}

To define CA \cite{Wolfram-2002,Ilachinski-2001} one should specify
\emph{i}) a regular grid $\mathcal{G}$ of sites $\xi$,
\emph{ii}) a set $\mathcal{S}=\{s_1,\cdots,s_N \}$ of available sites states
\emph{iii}) and a rule $\mathcal{F}$ governing the time evolution of the system.
The latter defines a state $s(\xi;t+1)$ of site $\xi\in\mathcal{G}$ at time $t+1$ basing of the states of site $\xi$ neighbourhood $\mathcal{N}$ at time $t$
\[
s(\xi;t+1)=\mathcal{F}\big(s(\xi;t),s(\xi_1;t),\cdots,s(\xi_M;t)\big)
\]
and $\xi_{i=1,\cdots,M}\in\mathcal{N}'=\mathcal{N}\setminus \{\xi\}$, where $M$ is a number of sites in deleted neighbourhood of $\xi$.

\subsection{Set $\mathcal{S}$}

Similarly to our earlier approaches~\cite{Kowalska-2017a-e,Kowalska-2017b-e} every agent is characterised by a Boolean vector variable 
\[
\mathbf{C}(\xi;t)=[c_1(\xi;t),c_2(\xi;t),\cdots,c_K(\xi,t)],
\]
where $c_i(\xi;t)\in\{0,1\}$ describes lack [$c_i(\xi;t)=0$] or possessing [$c_i(\xi;t)=1$] $i$-th chunk of knowledge by the agent at site $\xi$ and at time $t$.
$K$ stands for the number of \emph{all} chunks of knowledge available for every agent.

\subsection{Rule $\mathcal{F}$}

The rule $\mathcal{F}$ states, that during each simulation step $t$ each agent may receive \emph{single} chunk of knowledge from the randomly selected agent in his/her deleted neighbourhood $\mathcal{N}'$.
However, the sender of this information (placed at position $\xi'\in\mathcal{N}'$) is willing to share his/her knowledge \emph{only} when the recipient is smart enough.
Namely, the chunk of knowledge is transferred from sender (at site $\xi'$) to recipient (at site $\xi$) only when the difference in number of possessed chunk of knowledge among these two agents is \emph{exactly} equal to one:
\begin{subequations}
\label{eq:rule}
\begin{equation}
\label{eq:rule-a}
c_i(\xi;t+1)=1\iff c_i(\xi;t)=0 \wedge c_i(\xi';t)=1
\end{equation}
\begin{equation}
\label{eq:rule-b}
\wedge \left[\sum_{j=1}^{K} c_j(\xi';t)-\sum_{j=1}^{K} c_j(\xi;t)\right]=1.
\end{equation}
\end{subequations}

Such approach is not different from~\citet{Deffuant-2000} model of opinion dynamics, where opinion exchange among agents is possible only when sender and recipient have similar opinions~\cite{Hegselmann-2002,Malarz2006b,Kulakowski-2009,Gronek2011,Kulakowski2014} and consistent with empirical findings regarding knowledge transfer in organisation~\cite{Reagans-2003}.
Also simulations driven by the homophily principle assume that ‘agents are likely to exhibit strong preferences towards agents with which they are similar’ \cite{Hirshman-2011}.

\subsection{Grid $\mathcal{G}$}

We assume that agents occupy the nodes of a square lattice with linear size $L$:
\[
\mathcal{G}=\{(x,y): x, y\in\mathbb{N} \wedge 1\le x,y\le L \}.
\]
Additionally, the periodic boundary conditions are assumed.

In order to check the influence of the range of interaction on efficiency and effectiveness of knowledge transfer we consider three kinds of neighbourhoods $\mathcal{N}$.
The site $\xi\in\mathcal{G}$ and its $M=4$ nearest-neighbours constitute {\em von Neumann neighbourhood}
\[
\mathcal{V}=\{(x,y), (x,y\pm 1), (x\pm 1,y) \},
\]
while for {\em Moore neighbourhood} also four next-nearest neighbours are included ($M=8$)
\[
\mathcal{M}=\mathcal{V} \cup \{(x-1, y\pm 1), (x+1, y\pm 1) \}.
\]
Finally, we apply a {\em complex neighbourhood} with next-next-nearest neighbours ($M=12$)
\[
\mathcal{C}=\mathcal{M} \cup \{(x, y\pm 2), (x\pm 2, y) \}.
\]

\begin{figure*}
\centering
\includegraphics[width=.30\textwidth]{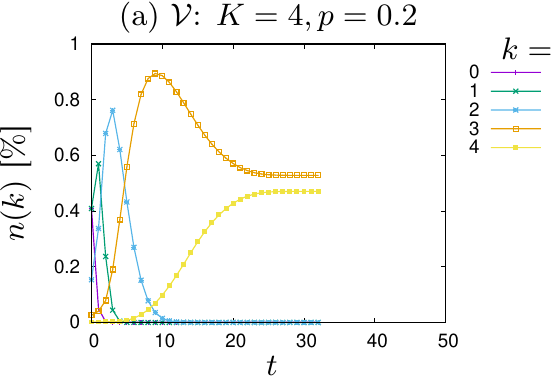}
\includegraphics[width=.30\textwidth]{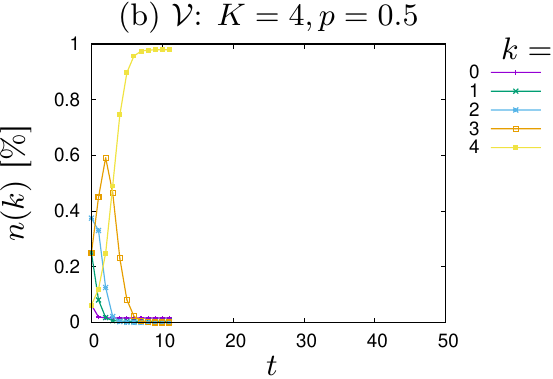}
\includegraphics[width=.30\textwidth]{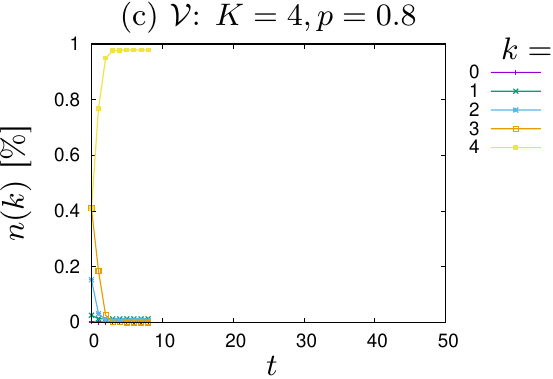}
\includegraphics[width=.30\textwidth]{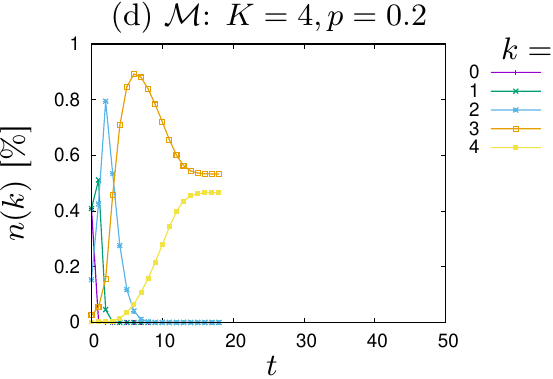}
\includegraphics[width=.30\textwidth]{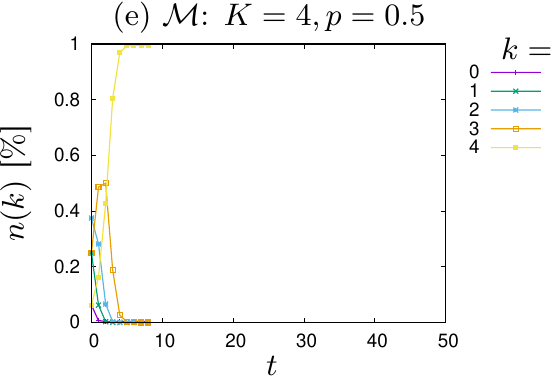}
\includegraphics[width=.30\textwidth]{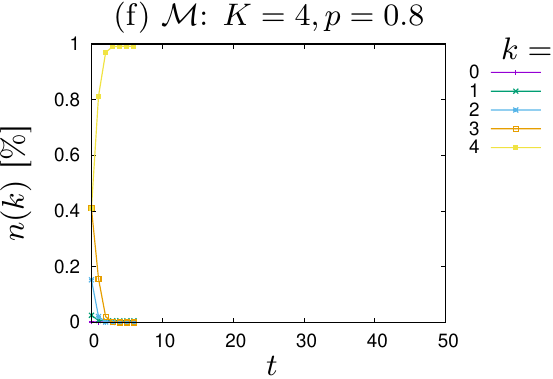}
\includegraphics[width=.30\textwidth]{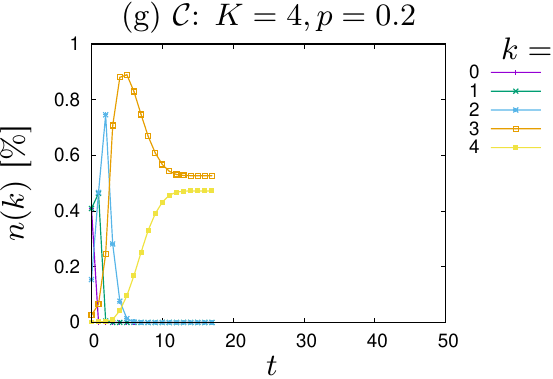}
\includegraphics[width=.30\textwidth]{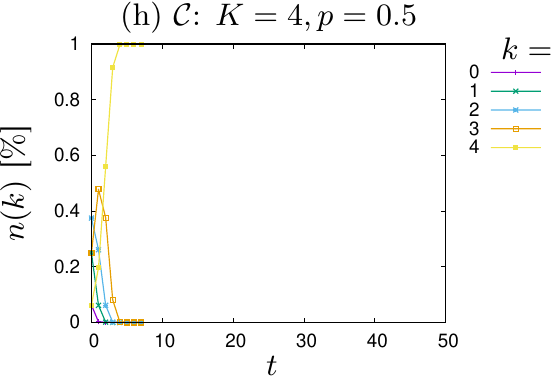}
\includegraphics[width=.30\textwidth]{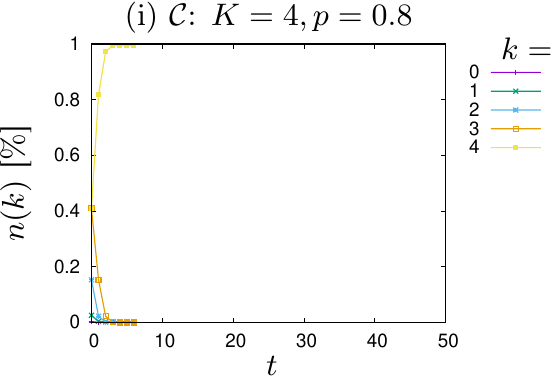}
\includegraphics[width=.30\textwidth]{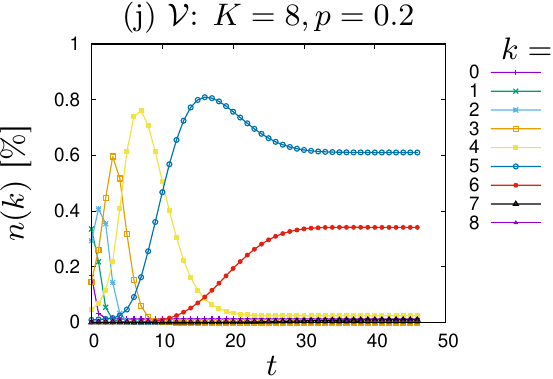}
\includegraphics[width=.30\textwidth]{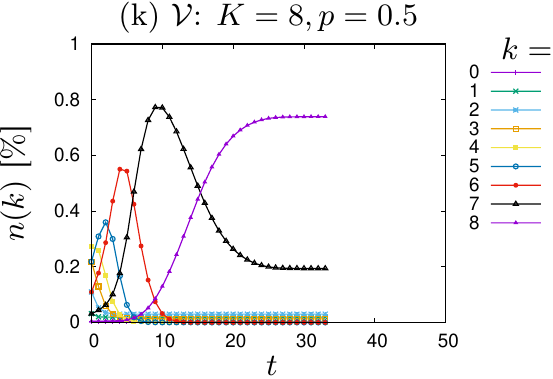}
\includegraphics[width=.30\textwidth]{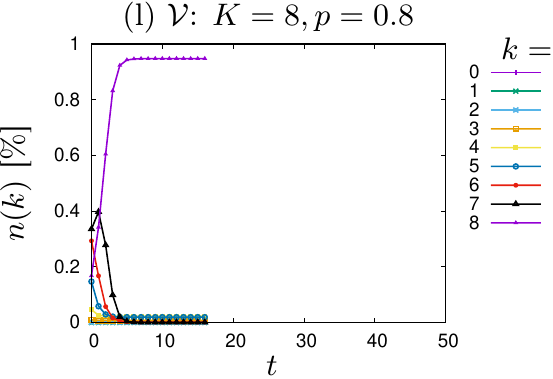}
\includegraphics[width=.30\textwidth]{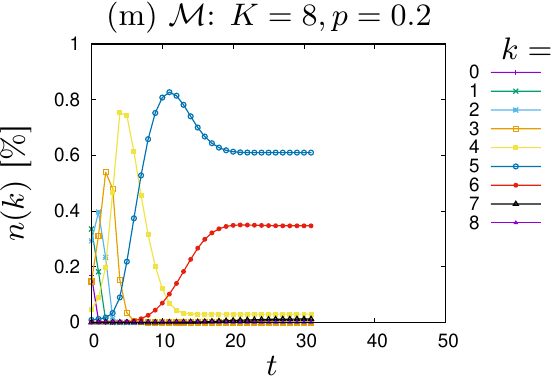}
\includegraphics[width=.30\textwidth]{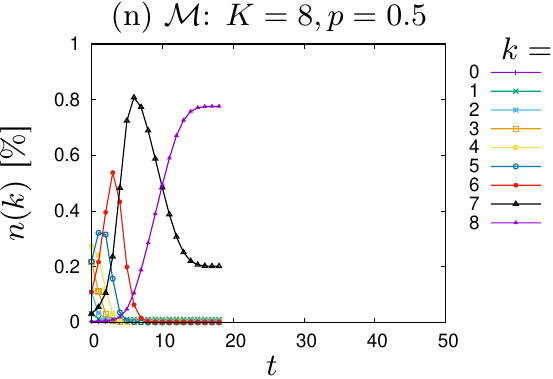}
\includegraphics[width=.30\textwidth]{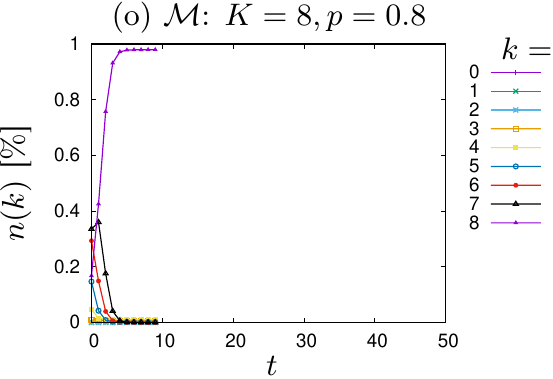}
\includegraphics[width=.30\textwidth]{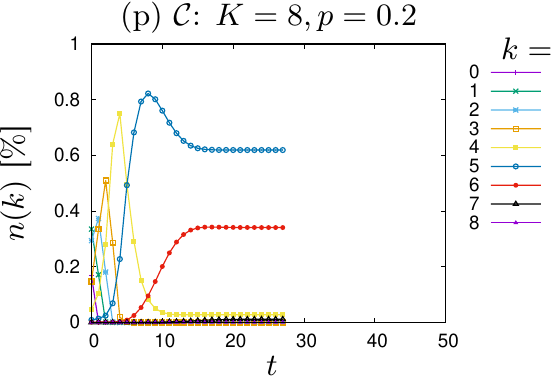}
\includegraphics[width=.30\textwidth]{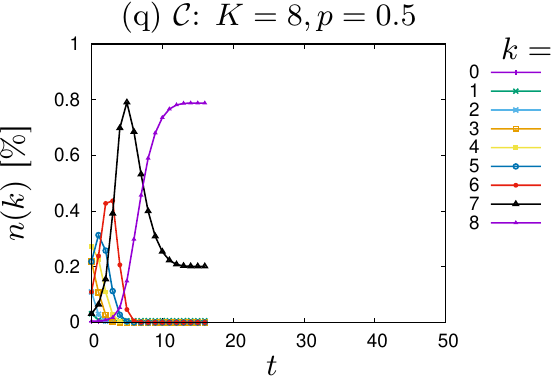}
\includegraphics[width=.30\textwidth]{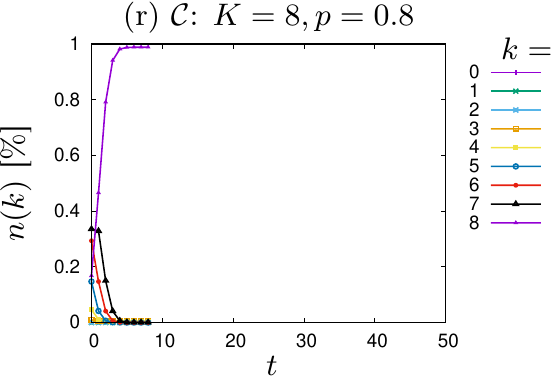}
\caption{\label{F:nk}The time evolution of the fraction $n(k)$ of agents having $k$ chunks of knowledge for $L=20$ and various \emph{i}) initial concentration of chunks of knowledge ($p=0.2$, 0.5, 0.8) and \emph{ii}) various values of $K$.
The results are obtained for von Neumann (a-c, j-l), Moore (d-f, m-o), and complex (g-i, p-r) neighbourhood.
The values of $n(k)$ are averaged over $R=10^4$ independent simulations.}
\end{figure*}

\section{\label{S:results}Results}

We measure the efficiency\footnote{\citet[p. 663]{Daft-1998} defines effectiveness as `the degree to which goals are attained' and efficiency as `amount of resources used to produce a unit of output'.} of the knowledge transfer as a time $\tau$ necessary for reaching the steady state of the system.
In order to evaluate the effectiveness of the knowledge transfer we study the average coverage of chunks of knowledge in the system $\langle f\rangle$ and the fraction $n(K)$ of agent having all ($K$) chunks of knowledge which are available in the system.

\begin{figure}
\centering
\includegraphics[width=.30\textwidth]{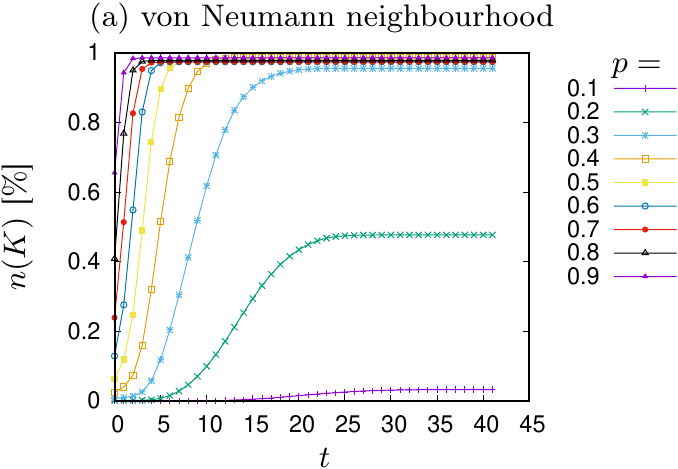}
\includegraphics[width=.30\textwidth]{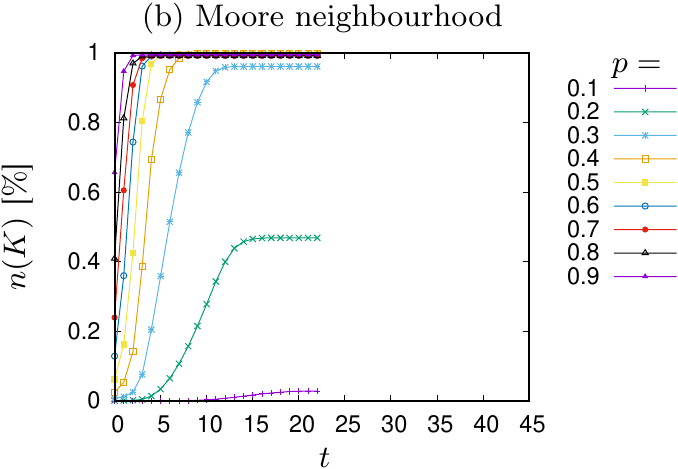}
\includegraphics[width=.30\textwidth]{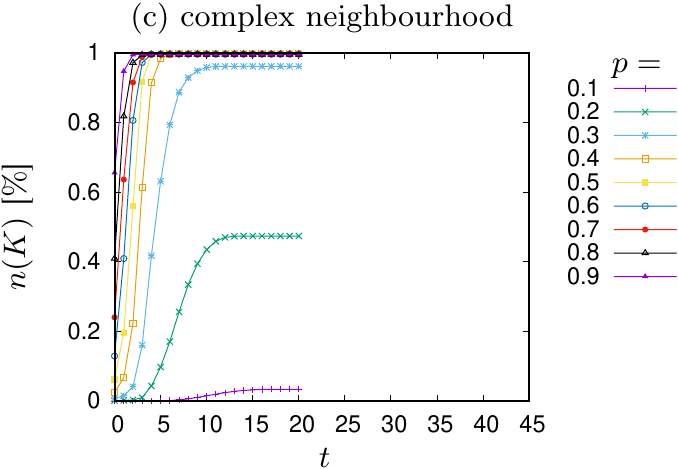}
\caption{\label{F:nK}The time evolution of the fraction $n(K)$ of agents having all available ($K$) chunks of knowledge for $L=20$, $K=4$ and various initial level of knowledge in organisation $p$.
The values of $n(K)$ are averaged over $R=10^4$ independent simulations.}
\end{figure}

In Fig.~\ref{F:nk} the time evolution of the fraction $n(k)$ of agents having $k$ chunks of knowledge for $L=20$ and various \emph{i}) initial concentration of chunks of knowledge ($p=0.2$, 0.5, 0.8) and \emph{ii}) various values of $K$ are presented.
The results are obtained for von Neumann (a-c, j-l), Moore (d-f, m-o), and complex (g-i, p-r) neighbourhood.
The values presented in Figs.~\ref{F:nk}-\ref{F:tau} are averaged over $R=10^4$ independent simulations.
As we can see, the shape of neighbourhood does not influence the time evolution of $n(k)$ too much.
The system is much more vulnerable to the changes of initial concentration of chunks of knowledge $p$ \cite{Kowalska-2017a-e}.
However---particularly for larger $K$ and larger $p$---we can see that fraction of $n(k=K)$ grows slightly with the number $M$ of sites constituting the neighbourhood.

\subsection{\label{S:effectiveness}Effectiveness of the knowledge transfer}

In Fig.~\ref{F:nK} the time evolution of the fraction $n(K)$ of agents having all available ($K$) chunks of knowledge for $L=20$, $K=4$ and various initial level of knowledge in organisation $p$ are presented.
Again, the shape of neighbourhood does not affect the level on which fraction $n(K)$ saturates, however, the time of reaching the stationary state is reduced roughly twice when we change von Neumann neighbourhood to Moore's one.
This may suggest that kind of neighbourhood may have greater impact on efficiency than on effectiveness of knowledge transfer.

\begin{figure*}
\centering
\includegraphics[width=.30\textwidth]{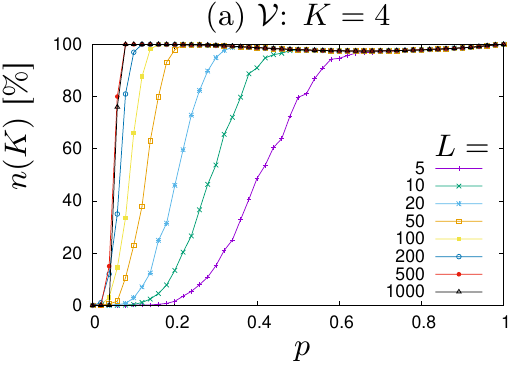}
\includegraphics[width=.30\textwidth]{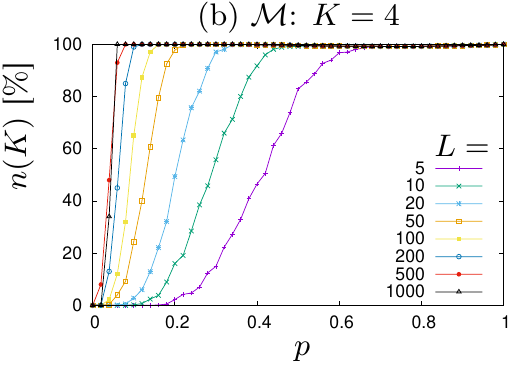}
\includegraphics[width=.30\textwidth]{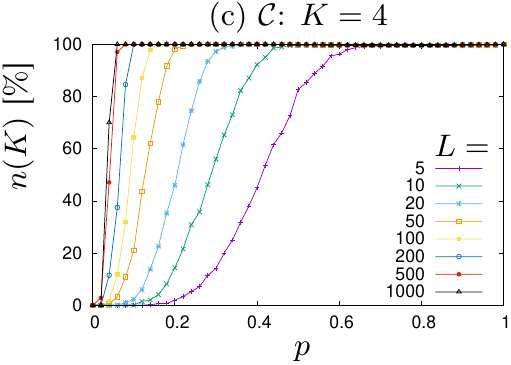}
\includegraphics[width=.30\textwidth]{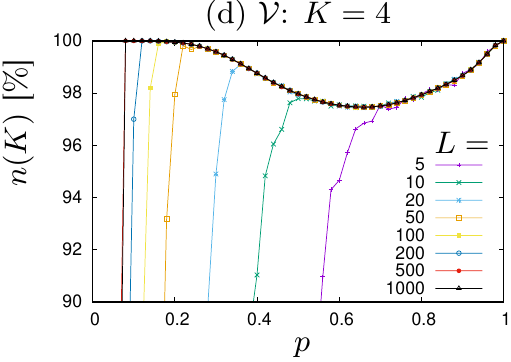}
\includegraphics[width=.30\textwidth]{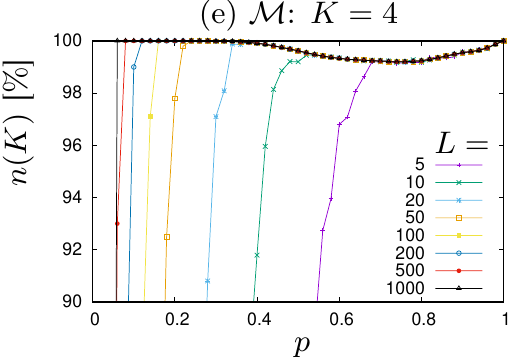}
\includegraphics[width=.30\textwidth]{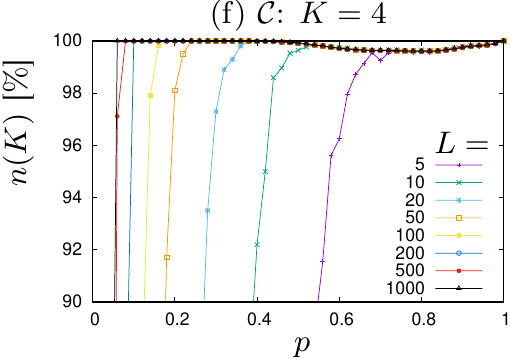}
\includegraphics[width=.30\textwidth]{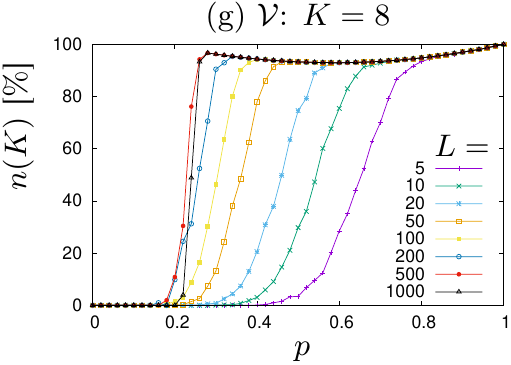}
\includegraphics[width=.30\textwidth]{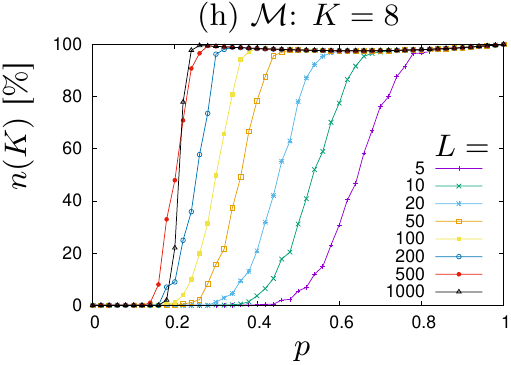}
\includegraphics[width=.30\textwidth]{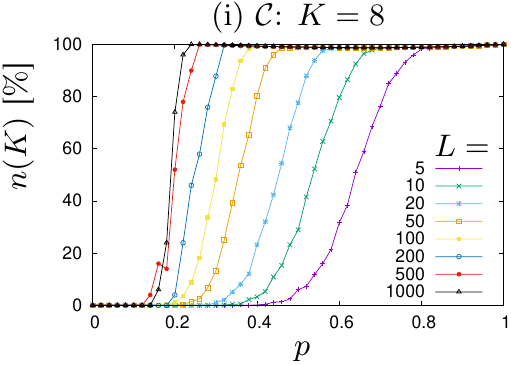}
\includegraphics[width=.30\textwidth]{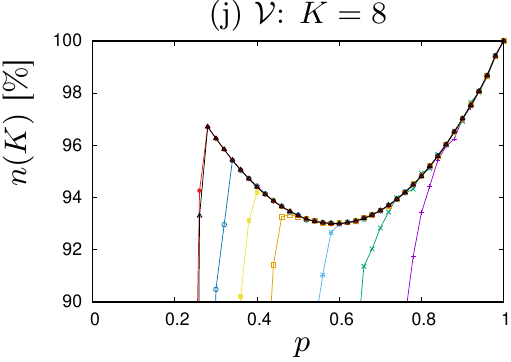}
\includegraphics[width=.30\textwidth]{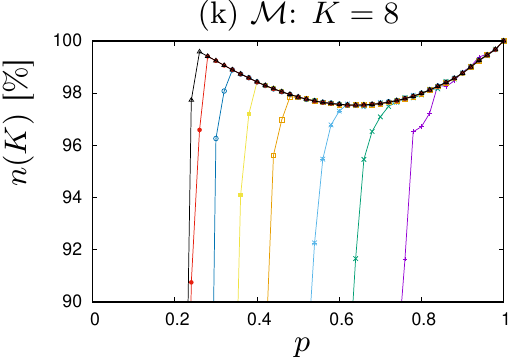}
\includegraphics[width=.30\textwidth]{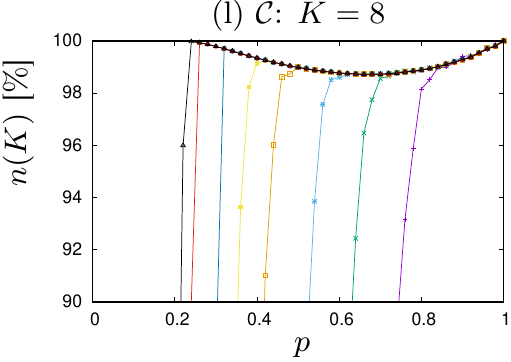}
\caption{\label{F:fvsp}The fraction of agents having total knowledge $\sum_{i=1}^K c_i=K=8$ vs. initial knowledge in the system ($p$) for various system sizes $L$ and various neighbourhoods.
The results are averaged over $R=10^4$ independent simulations.}
\end{figure*}

In Fig.~\ref{F:fvsp} the fraction of agents having total knowledge $\sum_{i=1}^K c_i=K$ vs. initial knowledge in the system ($p$) for various system sizes $L$ and various neighbourhoods are presented.
The figures~\ref{F:fvsp}(d, e, f) and~\ref{F:fvsp}(j, k, l) present top 10\% of figures~\ref{F:fvsp}(a, b, c) and~\ref{F:fvsp}(g, h, i), respectively.
For von Neumann neighbourhood and medium ($L=10$) and larger organisations we observe minimum of $n(K)$ curves for $p\approx 0.6$~\cite{Kowalska-2017a-e}.
This counter-intuitive effect is directly associated with restriction~\eqref{eq:rule-b}---for high enough initial concentration of chunks of knowledge $p$ some agents acquire high level of competences quite quickly and do not want share their knowledge with their not-so-smart neighbours.
The effect may be reduced~\cite{Kowalska-2017b-e} when agents receive chunks of knowledge from smarter agents
\begin{subequations}
\begin{equation}
\label{eq:rule-c}
\left[\sum_{j=1}^{K} c_j(\xi';t)-\sum_{j=1}^{K} c_j(\xi;t)\right]\ge 1
\end{equation}
and even vanished~\cite{Kowalska-2017b-e} when sender is smarter than or as smart as recipient of chunk of knowledge
\begin{equation}
\label{eq:rule-d}
\left[\sum_{j=1}^{K} c_j(\xi';t)-\sum_{j=1}^{K} c_j(\xi;t)\right]\ge 0.
\end{equation}
\end{subequations}
Also increasing the range of interaction may help in reducing this effect.
The reduction is stronger for complex neighbourhood [Fig.~\ref{F:fvsp}(c, f, i, l)] than for Moore's one [Fig.~\ref{F:fvsp}(b, e, h, k)].

\begin{figure*}
\centering
\includegraphics[width=.30\textwidth]{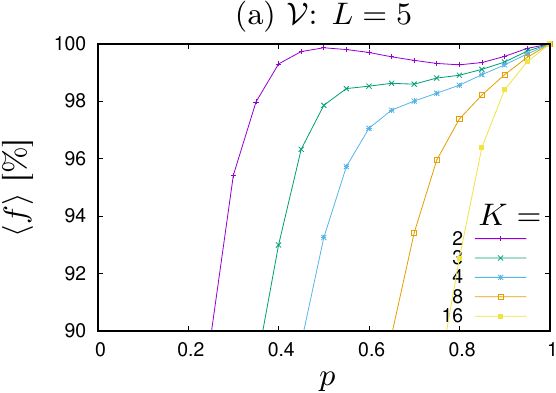}
\includegraphics[width=.30\textwidth]{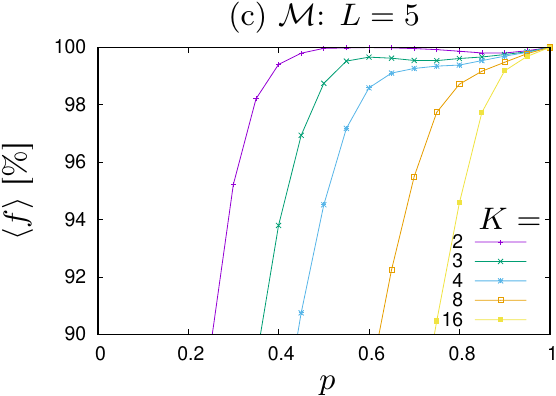}
\includegraphics[width=.30\textwidth]{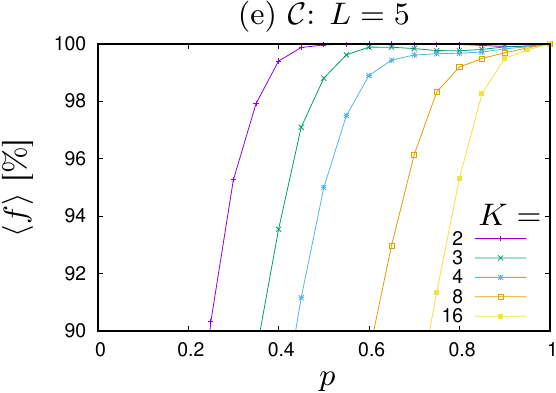}
\includegraphics[width=.30\textwidth]{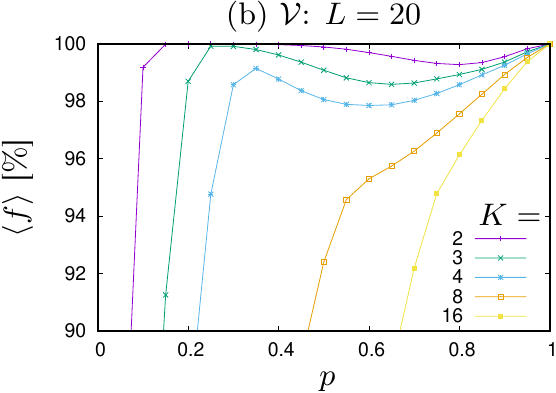}
\includegraphics[width=.30\textwidth]{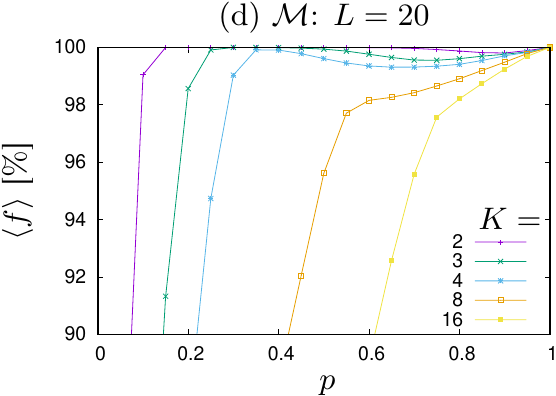}
\includegraphics[width=.30\textwidth]{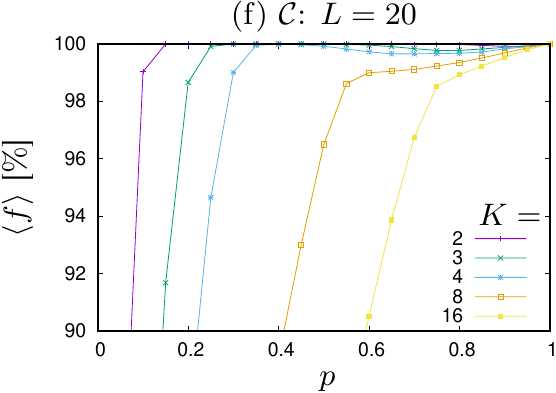}
\caption{\label{F:f}The average coverage $\langle f\rangle$ of chunks of knowledge in organisation for small (a, c, e, $L=5$) and average (b, d, f, $L=20$) size of organisation and various neighbourhoods.
The values of $\langle f\rangle$ are averaged over $R=10^4$ independent simulations.}
\end{figure*}

In Fig.~\ref{F:f} the average coverage $\langle f\rangle$ of chunks of knowledge in organisation for small [$L=5$, Figs.~\ref{F:f}(a, c, e)] and medium [$L=20$, Figs.~\ref{F:f}(b, d, f)] sizes of organisation and various values of $K$ and neighbourhoods are presented.
And again, for $K<8$ one may observe non-monotonous dependence of the average coverage $\langle f\rangle$ of chunks of knowledge in organisation vs. initial concentration of chunks of knowledge $p$ \cite{Kowalska-2017a-e}.
The changes in knowledge transfer rules from Eq.~\eqref{eq:rule-b} to Eq.~\eqref{eq:rule-c} or Eq.~\eqref{eq:rule-d} generates monotonous dependence $\langle f\rangle$ vs. $p$ \cite{Kowalska-2017b-e}. 
Moreover, for rule Eq.~\eqref{eq:rule-d} we observe collapse of curves for various values of $K$ to a single curve close to Heaviside's step function $\Theta(p-0.05)$. 
The changes in range of neighbourhoods do not yield so spectacular changes in shapes of $\langle f\rangle$ vs. $p$ curves.
However, the reduction of ineffectiveness of knowledge transfer for all considered values of $K$ may be observed as increasing minimum of $\langle f\rangle$ vs. $p$ dependencies in interval $p\in[0.5,1]$.

\subsection{Efficiency of the knowledge transfer}

\begin{figure*}
\centering
\includegraphics[width=.30\textwidth]{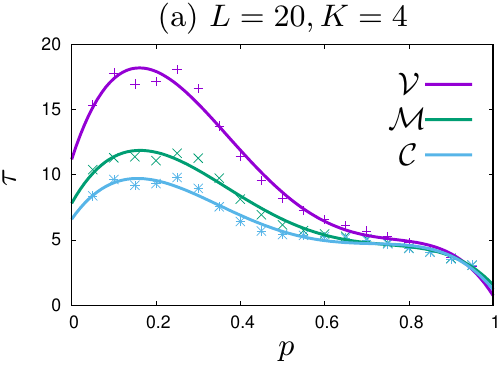}
\includegraphics[width=.30\textwidth]{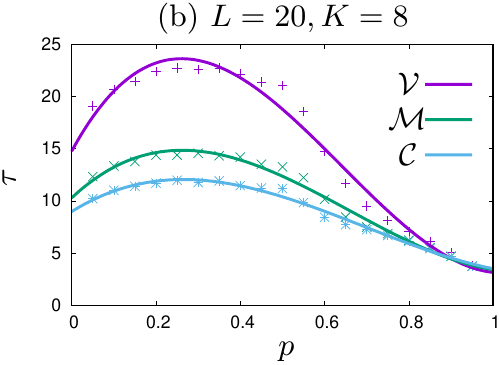}
\caption{\label{F:tau}The time $\tau$ necessary for reaching the stationary state of the system for (a) $K=4$ and (b) $K=8$ chunks of knowledge available in the system for $L=20$.
The values of $\tau$ are averaged over $R=10^4$ independent simulations.
The solid line curves show low degree polynomial fits.}
\end{figure*}

In Fig.~\ref{F:tau} the times $\tau$ necessary for reaching the stationary state of the system for (a) $K=4$ and (b) $K=8$ chunks of knowledge available in the system for $L=20$ are presented.
The solid line curves show low degree polynomial fits as a guide for eyes.
As we mentioned in Sec.~\ref{S:effectiveness} the number of sites in neighbourhood may influence the efficiency of knowledge transfer more than effectiveness of this process.
And indeed, the change of neighbourhood may lead to reduction of time $\tau$ of reaching the stationary state of the system even twice.
This difference vanishes for larger values of initial concentration of chunks of knowledge in organisation ($p>0.8$) as in this limit independently on kind of neighbourhood almost all agents acquires almost all chunks of knowledge in several time steps.

\section{\label{S:disc}Discussion and conclusions}
The proposed model has been designed to investigate the impact of the informal groups in the organization on the effectiveness and efficiency of knowledge transfer.
Three different neighbourhoods: von Neumann neighbourhood with four nearest-neighbours, Moore neighbourhood with eight neighbours and complex one with twelve neighbours have been adopted.
Our results show that the size of the neighbourhood has a far greater impact on the efficiency than the effectiveness of the knowledge transfer.
As can be seen, the knowledge transfer time is shorter in the case of a larger neighbourhood.
This may be related to the coherence of the network of agents.
Greater cohesion occurs when people have dense and overlapping relationships \cite{Fleming-2007}.
This situation of overlapping ties occurs in the Moore’s neighbourhood for eight neighbours and complex neighbourhood for twelve neighbours.
If the lattice is completely full, the agents have a greater number of common neighbours (and thus overlapping ties) than those of von Neumann’s neighbourhood for four nearest-neighbours.
In the context of social capital considerations, a closed social network (i.e., more coherent) raises greater trust between people and thus improves the flow of information \cite{Coleman-1988}.
This is also confirmed by the results of our simulation.

Of course, the results of the simulations encourage further research and exploration of the factors that influence the transfer of knowledge, as well as the premise of empirical research on the impact of employee-to-employee networking on knowledge exchange.
The model presented in this work can be easily extended and other factors describing the transfer of knowledge could be taken into account.
One of them could be e.g. homophily because research show that it affects the frequency of the interaction between agents \cite{Hirshman-2011}.  

\begin{acknowledgments}
This research was supported by \href{https://www.ncn.gov.pl/?language=en}{National Science Centre} (NCN) in Poland (grant no. UMO-2014/15/B/HS4/04433) and partially by Polish Ministry of Science and Higher Education.
\end{acknowledgments}

\bibliography{../../opus8}

\begin{thebibliography}{26}%
\makeatletter
\providecommand \@ifxundefined [1]{%
 \@ifx{#1\undefined}
}%
\providecommand \@ifnum [1]{%
 \ifnum #1\expandafter \@firstoftwo
 \else \expandafter \@secondoftwo
 \fi
}%
\providecommand \@ifx [1]{%
 \ifx #1\expandafter \@firstoftwo
 \else \expandafter \@secondoftwo
 \fi
}%
\providecommand \natexlab [1]{#1}%
\providecommand \enquote  [1]{``#1''}%
\providecommand \bibnamefont  [1]{#1}%
\providecommand \bibfnamefont [1]{#1}%
\providecommand \citenamefont [1]{#1}%
\providecommand \href@noop [0]{\@secondoftwo}%
\providecommand \href [0]{\begingroup \@sanitize@url \@href}%
\providecommand \@href[1]{\@@startlink{#1}\@@href}%
\providecommand \@@href[1]{\endgroup#1\@@endlink}%
\providecommand \@sanitize@url [0]{\catcode `\\12\catcode `\$12\catcode
  `\&12\catcode `\#12\catcode `\^12\catcode `\_12\catcode `\%12\relax}%
\providecommand \@@startlink[1]{}%
\providecommand \@@endlink[0]{}%
\providecommand \url  [0]{\begingroup\@sanitize@url \@url }%
\providecommand \@url [1]{\endgroup\@href {#1}{\urlprefix }}%
\providecommand \urlprefix  [0]{URL }%
\providecommand \Eprint [0]{\href }%
\providecommand \doibase [0]{http://dx.doi.org/}%
\providecommand \selectlanguage [0]{\@gobble}%
\providecommand \bibinfo  [0]{\@secondoftwo}%
\providecommand \bibfield  [0]{\@secondoftwo}%
\providecommand \translation [1]{[#1]}%
\providecommand \BibitemOpen [0]{}%
\providecommand \bibitemStop [0]{}%
\providecommand \bibitemNoStop [0]{.\EOS\space}%
\providecommand \EOS [0]{\spacefactor3000\relax}%
\providecommand \BibitemShut  [1]{\csname bibitem#1\endcsname}%
\let\auto@bib@innerbib\@empty
\bibitem [{\citenamefont {Chen}(2004)}]{Chen-2004}%
  \BibitemOpen
  \bibfield  {author} {\bibinfo {author} {\bibfnamefont {C.}~\bibnamefont
  {Chen}},\ }\bibfield  {title} {\enquote {\bibinfo {title} {The effects of
  knowledge attribute, alliance characteristics, and absorptive capacity on
  knowledge transfer performance},}\ }\href {\doibase
  10.1111/j.1467-9310.2004.00341.x} {\bibfield  {journal} {\bibinfo  {journal}
  {R\&D Management}\ }\textbf {\bibinfo {volume} {34}},\ \bibinfo {pages}
  {311--321} (\bibinfo {year} {2004})}\BibitemShut {NoStop}%
\bibitem [{\citenamefont {Lyles}\ and\ \citenamefont
  {Salk}(1996)}]{Lyles-1996}%
  \BibitemOpen
  \bibfield  {author} {\bibinfo {author} {\bibfnamefont {M.}~\bibnamefont
  {Lyles}}\ and\ \bibinfo {author} {\bibfnamefont {J.}~\bibnamefont {Salk}},\
  }\bibfield  {title} {\enquote {\bibinfo {title} {Knowledge acquisition from
  foreign parents in international joint venture: An empirical examination in
  the hungarian context},}\ }\href {\doibase 10.1057/palgrave.jibs.8400243}
  {\bibfield  {journal} {\bibinfo  {journal} {Journal of International Business
  Studies}\ }\textbf {\bibinfo {volume} {27}},\ \bibinfo {pages} {877--903}
  (\bibinfo {year} {1996})}\BibitemShut {NoStop}%
\bibitem [{\citenamefont {Tsai}(2001)}]{Tsai-2001}%
  \BibitemOpen
  \bibfield  {author} {\bibinfo {author} {\bibfnamefont {W.}~\bibnamefont
  {Tsai}},\ }\bibfield  {title} {\enquote {\bibinfo {title} {Knowledge transfer
  in intraorganizational networks: Effects of network position and absorptive
  capacity on business unit innovation and performance},}\ }\href {\doibase
  10.2307/3069443} {\bibfield  {journal} {\bibinfo  {journal} {Academy of
  Management Journal}\ }\textbf {\bibinfo {volume} {44}},\ \bibinfo {pages}
  {996--1004} (\bibinfo {year} {2001})}\BibitemShut {NoStop}%
\bibitem [{\citenamefont {Applehans}\ \emph {et~al.}(1999)\citenamefont
  {Applehans}, \citenamefont {Globe},\ and\ \citenamefont
  {Laugero}}]{Applehans-1999}%
  \BibitemOpen
  \bibfield  {author} {\bibinfo {author} {\bibfnamefont {W.}~\bibnamefont
  {Applehans}}, \bibinfo {author} {\bibfnamefont {A.}~\bibnamefont {Globe}}, \
  and\ \bibinfo {author} {\bibfnamefont {G.}~\bibnamefont {Laugero}},\
  }\href@noop {} {\emph {\bibinfo {title} {Managing knowledge. A practical
  web-based approach}}}\ (\bibinfo  {publisher} {Addison-Wesley},\ \bibinfo
  {address} {Reading, MA},\ \bibinfo {year} {1999})\BibitemShut {NoStop}%
\bibitem [{\citenamefont {Davenport}\ and\ \citenamefont
  {Prusak}(2000)}]{Davenport-2000}%
  \BibitemOpen
  \bibfield  {author} {\bibinfo {author} {\bibfnamefont {T.~H.}\ \bibnamefont
  {Davenport}}\ and\ \bibinfo {author} {\bibfnamefont {L.}~\bibnamefont
  {Prusak}},\ }\href@noop {} {\emph {\bibinfo {title} {Working knowledge: How
  organizations manage what they know}}}\ (\bibinfo  {publisher} {Harvard
  Business School Press},\ \bibinfo {address} {Boston, MA},\ \bibinfo {year}
  {2000})\BibitemShut {NoStop}%
\bibitem [{\citenamefont {Wah}(1999)}]{Wah-1999}%
  \BibitemOpen
  \bibfield  {author} {\bibinfo {author} {\bibfnamefont {L.}~\bibnamefont
  {Wah}},\ }\bibfield  {title} {\enquote {\bibinfo {title} {Behind the buzz},}\
  }\href@noop {} {\bibfield  {journal} {\bibinfo  {journal} {Management
  Review}\ }\textbf {\bibinfo {volume} {88}},\ \bibinfo {pages} {17--26}
  (\bibinfo {year} {1999})}\BibitemShut {NoStop}%
\bibitem [{\citenamefont {Argote}\ and\ \citenamefont
  {Ingram}(2000)}]{Argote-2000}%
  \BibitemOpen
  \bibfield  {author} {\bibinfo {author} {\bibfnamefont {L.}~\bibnamefont
  {Argote}}\ and\ \bibinfo {author} {\bibfnamefont {P.}~\bibnamefont
  {Ingram}},\ }\bibfield  {title} {\enquote {\bibinfo {title} {Knowledge
  transfer: a basis for competitive advantage in firms},}\ }\href {\doibase
  10.1006/obhd.2000.2893} {\bibfield  {journal} {\bibinfo  {journal}
  {Organizational Behavior and Human Decision Processes}\ }\textbf {\bibinfo
  {volume} {82}},\ \bibinfo {pages} {150–169} (\bibinfo {year}
  {2000})}\BibitemShut {NoStop}%
\bibitem [{\citenamefont {Reagans}\ and\ \citenamefont
  {McEvily}(2003)}]{Reagans-2003}%
  \BibitemOpen
  \bibfield  {author} {\bibinfo {author} {\bibfnamefont {R.}~\bibnamefont
  {Reagans}}\ and\ \bibinfo {author} {\bibfnamefont {B.}~\bibnamefont
  {McEvily}},\ }\bibfield  {title} {\enquote {\bibinfo {title} {Network
  structure and knowledge transfer: The effects of cohesion and range},}\
  }\href {\doibase 10.2307/3556658} {\bibfield  {journal} {\bibinfo  {journal}
  {Administrative Science Quarterly}\ }\textbf {\bibinfo {volume} {48}},\
  \bibinfo {pages} {240--267} (\bibinfo {year} {2003})}\BibitemShut {NoStop}%
\bibitem [{\citenamefont {Ingram}\ and\ \citenamefont
  {Roberts}(2000)}]{Ingram-2000}%
  \BibitemOpen
  \bibfield  {author} {\bibinfo {author} {\bibfnamefont {P.}~\bibnamefont
  {Ingram}}\ and\ \bibinfo {author} {\bibfnamefont {P.}~\bibnamefont
  {Roberts}},\ }\bibfield  {title} {\enquote {\bibinfo {title} {Friendships
  among competitors in the {S}ydney hotel industry},}\ }\href {\doibase
  10.1086/316965} {\bibfield  {journal} {\bibinfo  {journal} {American Journal
  of Sociology}\ }\textbf {\bibinfo {volume} {106}},\ \bibinfo {pages}
  {387--423} (\bibinfo {year} {2000})}\BibitemShut {NoStop}%
\bibitem [{\citenamefont {Reagan}\ and\ \citenamefont
  {Zuckerman}(2001)}]{Reagans-2001}%
  \BibitemOpen
  \bibfield  {author} {\bibinfo {author} {\bibfnamefont {R.}~\bibnamefont
  {Reagan}}\ and\ \bibinfo {author} {\bibfnamefont {E.~W.}\ \bibnamefont
  {Zuckerman}},\ }\bibfield  {title} {\enquote {\bibinfo {title} {Networks,
  diversity, and productivity: {T}he social capital of corporate {R}\&{D}
  teams},}\ }\href {\doibase 10.1287/orsc.12.4.502.10637} {\bibfield  {journal}
  {\bibinfo  {journal} {Organization Science}\ }\textbf {\bibinfo {volume}
  {12}},\ \bibinfo {pages} {502–517} (\bibinfo {year} {2001})}\BibitemShut
  {NoStop}%
\bibitem [{\citenamefont {Chen}\ and\ \citenamefont {Huang}(2007)}]{Chen-2007}%
  \BibitemOpen
  \bibfield  {author} {\bibinfo {author} {\bibfnamefont {C.}~\bibnamefont
  {Chen}}\ and\ \bibinfo {author} {\bibfnamefont {J.}~\bibnamefont {Huang}},\
  }\bibfield  {title} {\enquote {\bibinfo {title} {How organizational climate
  and structure affect knowledge management---the social interaction
  perspective},}\ }\href {\doibase 10.1016/j.ijinfomgt.2006.11.001} {\bibfield
  {journal} {\bibinfo  {journal} {International Journal of Information
  Management}\ }\textbf {\bibinfo {volume} {27}},\ \bibinfo {pages} {104--118}
  (\bibinfo {year} {2007})}\BibitemShut {NoStop}%
\bibitem [{\citenamefont {Girdauskiene}\ and\ \citenamefont
  {Savaneviciene}(2012)}]{Girdauskiene-2012}%
  \BibitemOpen
  \bibfield  {author} {\bibinfo {author} {\bibfnamefont {L.}~\bibnamefont
  {Girdauskiene}}\ and\ \bibinfo {author} {\bibfnamefont {A.}~\bibnamefont
  {Savaneviciene}},\ }\bibfield  {title} {\enquote {\bibinfo {title}
  {Leadership role implementing knowledge transfer in creative organization:
  how does it work?}}\ }\href {\doibase 10.1016/j.sbspro.2012.04.002}
  {\bibfield  {journal} {\bibinfo  {journal} {Procedia---Social and Behavioral
  Sciences}\ }\textbf {\bibinfo {volume} {41}},\ \bibinfo {pages} {15--22}
  (\bibinfo {year} {2012})}\BibitemShut {NoStop}%
\bibitem [{\citenamefont {Kowalska-Stycze\'n}\ \emph
  {et~al.}(2017{\natexlab{a}})\citenamefont {Kowalska-Stycze\'n}, \citenamefont
  {Malarz},\ and\ \citenamefont {Paradowski}}]{Kowalska-2017a-e}%
  \BibitemOpen
  \bibfield  {author} {\bibinfo {author} {\bibfnamefont {A.}~\bibnamefont
  {Kowalska-Stycze\'n}}, \bibinfo {author} {\bibfnamefont {K.}~\bibnamefont
  {Malarz}}, \ and\ \bibinfo {author} {\bibfnamefont {K.}~\bibnamefont
  {Paradowski}},\ }\href@noop {} {\enquote {\bibinfo {title} {Model of
  knowledge transfer within an organization},}\ } (\bibinfo {year}
  {2017}{\natexlab{a}}),\ \Eprint {http://arxiv.org/abs/1704.07589
  [physics.soc-ph]} {arXiv:1704.07589 [physics.soc-ph]} \BibitemShut {NoStop}%
\bibitem [{\citenamefont {Kowalska-Stycze\'n}\ \emph
  {et~al.}(2017{\natexlab{b}})\citenamefont {Kowalska-Stycze\'n}, \citenamefont
  {Malarz},\ and\ \citenamefont {Paradowski}}]{Kowalska-2017b-e}%
  \BibitemOpen
  \bibfield  {author} {\bibinfo {author} {\bibfnamefont {A.}~\bibnamefont
  {Kowalska-Stycze\'n}}, \bibinfo {author} {\bibfnamefont {K.}~\bibnamefont
  {Malarz}}, \ and\ \bibinfo {author} {\bibfnamefont {K.}~\bibnamefont
  {Paradowski}},\ }\href@noop {} {\enquote {\bibinfo {title} {Searching for
  effective and efficient way of~knowledge transfer within an organization},}\
  } (\bibinfo {year} {2017}{\natexlab{b}}),\ \Eprint
  {http://arxiv.org/abs/1710.07924 [physics.soc-ph]} {arXiv:1710.07924
  [physics.soc-ph]} \BibitemShut {NoStop}%
\bibitem [{\citenamefont {Wolfram}(2002)}]{Wolfram-2002}%
  \BibitemOpen
  \bibfield  {author} {\bibinfo {author} {\bibfnamefont {S.}~\bibnamefont
  {Wolfram}},\ }\href {http://www.wolframscience.com/} {\emph {\bibinfo {title}
  {A new kind of science}}}\ (\bibinfo  {publisher} {Wolfram Media},\ \bibinfo
  {year} {2002})\BibitemShut {NoStop}%
\bibitem [{\citenamefont {Ilachinski}(2001)}]{Ilachinski-2001}%
  \BibitemOpen
  \bibfield  {author} {\bibinfo {author} {\bibfnamefont {A.}~\bibnamefont
  {Ilachinski}},\ }\href {\doibase 10.1142/4702} {\emph {\bibinfo {title}
  {Cellular automata: A discrete Universe}}}\ (\bibinfo  {publisher} {World
  Scientific},\ \bibinfo {year} {2001})\BibitemShut {NoStop}%
\bibitem [{\citenamefont {Deffuant}\ \emph {et~al.}(2000)\citenamefont
  {Deffuant}, \citenamefont {Neau}, \citenamefont {Amblard},\ and\
  \citenamefont {Weisbuch}}]{Deffuant-2000}%
  \BibitemOpen
  \bibfield  {author} {\bibinfo {author} {\bibfnamefont {G.}~\bibnamefont
  {Deffuant}}, \bibinfo {author} {\bibfnamefont {D.}~\bibnamefont {Neau}},
  \bibinfo {author} {\bibfnamefont {F.}~\bibnamefont {Amblard}}, \ and\
  \bibinfo {author} {\bibfnamefont {G.}~\bibnamefont {Weisbuch}},\ }\bibfield
  {title} {\enquote {\bibinfo {title} {Mixing beliefs among interacting
  agents},}\ }\href {\doibase 10.1142/S0219525900000078} {\bibfield  {journal}
  {\bibinfo  {journal} {Advances in Complex Systems}\ }\textbf {\bibinfo
  {volume} {3}},\ \bibinfo {pages} {87} (\bibinfo {year} {2000})}\BibitemShut
  {NoStop}%
\bibitem [{\citenamefont {Hegselmann}\ and\ \citenamefont
  {Krause}(2002)}]{Hegselmann-2002}%
  \BibitemOpen
  \bibfield  {author} {\bibinfo {author} {\bibfnamefont {R.}~\bibnamefont
  {Hegselmann}}\ and\ \bibinfo {author} {\bibfnamefont {U.}~\bibnamefont
  {Krause}},\ }\bibfield  {title} {\enquote {\bibinfo {title} {Opinion dynamics
  and bounded confidence: models, analysis and simulation},}\ }\href
  {http://jasss.soc.surrey.ac.uk/5/3/2.html} {\bibfield  {journal} {\bibinfo
  {journal} {JASSS---the Journal of Artificial Societies and Social
  Simulation}\ }\textbf {\bibinfo {volume} {5}},\ \bibinfo {pages} {2}
  (\bibinfo {year} {2002})}\BibitemShut {NoStop}%
\bibitem [{\citenamefont {Malarz}(2006)}]{Malarz2006b}%
  \BibitemOpen
  \bibfield  {author} {\bibinfo {author} {\bibfnamefont {K.}~\bibnamefont
  {Malarz}},\ }\bibfield  {title} {\enquote {\bibinfo {title} {Truth seekers in
  opinion dynamics models},}\ }\href {\doibase 10.1142/S0129183106009850}
  {\bibfield  {journal} {\bibinfo  {journal} {International Journal of Modern
  Physics C}\ }\textbf {\bibinfo {volume} {17}},\ \bibinfo {pages} {1521--1524}
  (\bibinfo {year} {2006})}\BibitemShut {NoStop}%
\bibitem [{\citenamefont {Ku{\l}akowski}(2009)}]{Kulakowski-2009}%
  \BibitemOpen
  \bibfield  {author} {\bibinfo {author} {\bibfnamefont {K.}~\bibnamefont
  {Ku{\l}akowski}},\ }\bibfield  {title} {\enquote {\bibinfo {title} {Opinion
  polarization in the receipt–accept–sample model},}\ }\href {\doibase
  10.1016/j.physa.2008.10.037} {\bibfield  {journal} {\bibinfo  {journal}
  {Physica A: Statistical Mechanics and its Applications}\ }\textbf {\bibinfo
  {volume} {388}},\ \bibinfo {pages} {469--476} (\bibinfo {year}
  {2009})}\BibitemShut {NoStop}%
\bibitem [{\citenamefont {Malarz}\ \emph {et~al.}(2011)\citenamefont {Malarz},
  \citenamefont {Gronek},\ and\ \citenamefont {Ku{\l}akowski}}]{Gronek2011}%
  \BibitemOpen
  \bibfield  {author} {\bibinfo {author} {\bibfnamefont {K.}~\bibnamefont
  {Malarz}}, \bibinfo {author} {\bibfnamefont {P.}~\bibnamefont {Gronek}}, \
  and\ \bibinfo {author} {\bibfnamefont {K.}~\bibnamefont {Ku{\l}akowski}},\
  }\bibfield  {title} {\enquote {\bibinfo {title} {{Z}aller--{D}effuant model
  of mass opinion},}\ }\href {\doibase 10.18564/jasss.1719} {\bibfield
  {journal} {\bibinfo  {journal} {JASSS---the Journal of Artificial Societies
  and Social Simulation}\ }\textbf {\bibinfo {volume} {14}},\ \bibinfo {pages}
  {2} (\bibinfo {year} {2011})}\BibitemShut {NoStop}%
\bibitem [{\citenamefont {Malarz}\ and\ \citenamefont
  {Ku{\l}akowski}(2014)}]{Kulakowski2014}%
  \BibitemOpen
  \bibfield  {author} {\bibinfo {author} {\bibfnamefont {K.}~\bibnamefont
  {Malarz}}\ and\ \bibinfo {author} {\bibfnamefont {K.}~\bibnamefont
  {Ku{\l}akowski}},\ }\bibfield  {title} {\enquote {\bibinfo {title} {Mental
  ability and common sense in an artificial society},}\ }\href {\doibase
  10.1051/epn/2014402} {\bibfield  {journal} {\bibinfo  {journal} {Europhysics
  News}\ }\textbf {\bibinfo {volume} {45}},\ \bibinfo {pages} {21--23}
  (\bibinfo {year} {2014})}\BibitemShut {NoStop}%
\bibitem [{\citenamefont {Hirshman}\ \emph {et~al.}(2011)\citenamefont
  {Hirshman}, \citenamefont {Charles},\ and\ \citenamefont
  {M.}}]{Hirshman-2011}%
  \BibitemOpen
  \bibfield  {author} {\bibinfo {author} {\bibfnamefont {B.~R.}\ \bibnamefont
  {Hirshman}}, \bibinfo {author} {\bibfnamefont {J.~S.}\ \bibnamefont
  {Charles}}, \ and\ \bibinfo {author} {\bibfnamefont {Carley~K.}\ \bibnamefont
  {M.}},\ }\bibfield  {title} {\enquote {\bibinfo {title} {Leaving us in tiers:
  can homophily be used to generate tiering effects?}}\ }\href {\doibase
  10.1007/s10588-011-9088-4} {\bibfield  {journal} {\bibinfo  {journal}
  {Computational and Mathematical Organization Theory}\ }\textbf {\bibinfo
  {volume} {17}},\ \bibinfo {pages} {318--343} (\bibinfo {year}
  {2011})}\BibitemShut {NoStop}%
\bibitem [{\citenamefont {Daft}(1998)}]{Daft-1998}%
  \BibitemOpen
  \bibfield  {author} {\bibinfo {author} {\bibfnamefont {R.~L.}\ \bibnamefont
  {Daft}},\ }\href@noop {} {\emph {\bibinfo {title} {Organization Theory and
  Design}}}\ (\bibinfo  {publisher} {South-Western College Publishing},\
  \bibinfo {address} {Cincinnati, OH},\ \bibinfo {year} {1998})\BibitemShut
  {NoStop}%
\bibitem [{\citenamefont {Fleming}\ \emph {et~al.}(2007)\citenamefont
  {Fleming}, \citenamefont {Mingo},\ and\ \citenamefont {Chen}}]{Fleming-2007}%
  \BibitemOpen
  \bibfield  {author} {\bibinfo {author} {\bibfnamefont {L.}~\bibnamefont
  {Fleming}}, \bibinfo {author} {\bibfnamefont {S.}~\bibnamefont {Mingo}}, \
  and\ \bibinfo {author} {\bibfnamefont {D.}~\bibnamefont {Chen}},\ }\bibfield
  {title} {\enquote {\bibinfo {title} {Collaborative brokerage, generative
  creativity, and creative success},}\ }\href {\doibase 10.2189/asqu.52.3.443}
  {\bibfield  {journal} {\bibinfo  {journal} {Administrative Science
  Quarterly}\ }\textbf {\bibinfo {volume} {52}},\ \bibinfo {pages} {443--475}
  (\bibinfo {year} {2007})}\BibitemShut {NoStop}%
\bibitem [{\citenamefont {Coleman}(1988)}]{Coleman-1988}%
  \BibitemOpen
  \bibfield  {author} {\bibinfo {author} {\bibfnamefont {J.~S.}\ \bibnamefont
  {Coleman}},\ }\bibfield  {title} {\enquote {\bibinfo {title} {Social capital
  in the creation of human capital},}\ }\href {\doibase 10.2307/2780243}
  {\bibfield  {journal} {\bibinfo  {journal} {American Journal of Sociology}\
  }\textbf {\bibinfo {volume} {94}},\ \bibinfo {pages} {S95--S120} (\bibinfo
  {year} {1988})}\BibitemShut {NoStop}%
\end{thebibliography}%
\end{document}